\documentclass[a4,11pt]{article}
\usepackage{amssymb}
\usepackage{amsmath}
\usepackage{epsfig}
\begin{document}
\newcommand{\cts}{\cos{\theta^*}}
\newcommand{\ct}{\cos{\theta}}
\newcommand{\sts}{\sin{\theta^*}}
\newcommand{\st}{\sin{\theta}}
\newcommand{\demi}{\frac{1}{2}}
\title{Detecting warm DM in the MeV/$c^2$ range}
\author{Fran\c{c}ois Vannucci \thanks{Laboratoire de Physique Nucl\'eaire et de Hautes Energies, CNRS - IN2P3 - Universit\'es Paris VI et Paris VII, Paris,France} \and Jean-Michel Levy \thanks{Idem} }
\sloppy
\maketitle
\begin{abstract}
Some tension exists between present experimental data and models which comprise only the three light neutrino mass eigenstates necessary to explain 
solar and atmospheric oscillation results. Hence the revival of the idea that additional more massive states might enter the active neutrino 
superpositions produced and observed in charged current reactions. Such 'heavy' neutrinos with masses in the keV or MeV range might also be of 
interest as dark matter candidates \cite{Shapo}. A state with mass larger than 1022 keV could decay into an $e^+ e^-$ pair and a light mass state, leaving 
an easily recognizable signature. The aim of this paper is to estimate the possible signal rate.
\end{abstract}
\pagenumbering{arabic}
\section{Massive "sterile" neutrinos}
Experimental results on neutrino oscillations demonstrate that there exist at least two oscillation lengths, hence two different squared mass 
differences at the level of and below $2.3\;10^{-3}\; \text{eV}^2$ . Together with the constraints on the absolute mass scale from $\nu_e$ mass measurements, 
these imply at least three states with masses below the eV scale. Moreover, $Z^0$ invisible width measurements at colliders indicate that there 
are no more than three light states with standard model (SM) couplings. However, this does not preclude the existence of more than 3 mass eigenstates 
since, for general Dirac and Majorana mass matrices, when SM active states are resolved into mass eigenstates, the couplings  with the $Z^0$ are reduced 
w.r.t. the SM value by a projection matrix which is a function of the neutrino mixing matrices elements. \\ Therefore here, we will hypothetize an 
additional, more massive state $N$, decoupled from the $Z^0$ mediated weak interactions, or at least from a possible non diagonal piece of the corresponding
currents, but still taking part in the mixing and therefore in the CC interactions with a strength given by the standard theory times $|U_{eN}|^2$, where 
$U_{eN}$ is the coefficient of $N$ in the linear combination defining $\nu_e$. Since 
radiative decays are very much depressed by loop propagators, and since a fast three light neutrino decay is forbidden by the non-diagonal $Z^0$ 
vanishing couplings, the first open decay channel for such a state of mass above 1 MeV would be $e^+ e^- \nu$. Provided $|U_{eN}|^2$ is sufficiently
small, this would give the $N$ state a long lifetime on the scale of the age of the Universe, making it a possible warm dark matter candidate which could 
be detectable by present means as we now show.

\section{Decay width}
Here we assume CC standard model couplings augmented by mixing matrix elements and no neutral current couplings. In this case, the heavy state decay
width into $e^- e^+ \nu$ is given by:
$$\Gamma = \frac{G^2 m^5}{96 \pi^3}|U_{eN}U_{el}|^2f(r)$$ where $U_{el}$ is the coefficient of some light neutrino mass eigenstate in the expansion of 
$\nu_e$ on the massive states basis \footnote{$|U_{el}|$ will henceforth be approximated by $1$ and the mass of the $l$ state will be neglected}, $G$
is Fermi's constant, $r = \frac{4m_e^2}{m^2}$ with $m_e$ the electron mass and $m$ is the mass of the $N$ state. The function $f$ can be easily calculated  
in the $V-A$ theory as (see Fig. 1): 
$$f(r) = 3r^2(1-\frac{r^2}{16})\text{argtanh}\sqrt{1-r} + (1-\frac{7}{2}r-\frac{1}{8}r^2-\frac{3}{16}r^3)\sqrt{1-r}$$
Note that in the $m_e \rightarrow 0$ and $U_{eN} \rightarrow 1$ limit and with the replacement $m \rightarrow m_{\mu}$ , $\Gamma$ is twice the standard 
muon decay width because we considered here that the $N$ state was of the Majorana type.\\
The CERN PS191 experiment \cite{PS191} gives an upper limit of $10^{-7}$ on $|U_{eN}|^2$ for $m = 100\; \text{MeV}$ whilst Borexino \cite{Borex} constrains 
it to be smaller than $10^{-3}\;-\;10^{-5}$ in the $1 - 8\;\text{MeV}$ mass range.

\section{Cosmological connection}
Dark matter (DM) amounts to $25\%$ of the Universe mass-energy density, whilst baryonic matter (BM) accounts only for $4\%$. On the other hand, the
ratio $n_B/n_{\gamma}$ of the numbers of baryons to photons is $6.10^{-10}$ from primordial nucleosynthesis. 
If we assume that DM is composed of $N$ particles, $n_N/n_{\gamma} \approx 25/4\cdot 6.\;10^{-10}$ if $N$ had the same mass as a baryon, or
$n_N/n_{\gamma} = 3.5\;10^{-6}/m$ for arbitrary $m$ in MeV \footnote{from now on, the symbol $m$ will stand for the mass of the $N$ state expressed in 
MeV}. Since $n_{\gamma} \approx 400 /\text{cm}^3$ and since each species of active relic neutrinos
amounts to $n_{\nu} \approx 100/\text{cm}^3$, we find $n_N/n_{\nu_e} \approx 1.4\;10^{-5}/m$\\
If all neutrino species are equally produced at high temperature, the initial $n_N/n_{\nu_e}$ ratio is of order $1$. Its present value results from the 
decay of $N$ over the life of the Universe, $T = 4.4\;10^{17} \text{s}$ \\ 
If $\tau$ is the lifetime of $N$ we therefore get: $e^{-T/\tau} = 1.4\;10^{-5}/m$ yielding
$\tau = 3.90\cdot 10^{16} \text{s},\;3.26\cdot 10^{16} \text{s},\;2.79\cdot 10^{16} \text{s}$  for $m = 1.1, \;10, \;100 \;\text{MeV}$ respectively.\\
Assuming that the $e^- e^+ \nu $ channel dominates the $N$ lifetime, one can then determine the corresponding $|U_{eN}|^2$ values through 
$m^5 |U|^2 = \frac{1.4\cdot 10^4}{\tau f(4m_e^2/m)}$, yielding $m^5|U|^2 = 4.14\cdot 10^{-10},\; 4.47\cdot 10^{-13},\; 5.01\cdot10^{-13}$ for the same 
three values of $m$ and it is readily seen that the mixing probed here is way lower than the present experimental limits mentioned above.

\section{Proposed detection method}
Galaxy rotation curves show that the missing mass corresponds to a local density of $300\;\text{MeV}/\text{cm}^3$ at the solar system \cite{DM}. Assuming 
this mass is made of $N$ particles with mass $m$ the Earth is travelling through a halo of numerical density $\rho = 300/m\; \text{cm}^{-3}$. Relative 
velocities are low and relativistic effects are completely negligible \footnote{The solar system velocity through the supposedly stationnary halo is about 
$200\; \text{km}/\text{s}$} and the thermal velocity of $2^{\circ}\;\text{K}$ neutrinos of mass 1 MeV is below 7 km/s. Therefore, a detector of volume $V$ 
contains $V\rho$ particles which have a decay probability $1/\tau \;\text{s}^{-1}$ yielding a potential $V\rho/\tau$ signal events per second. 
For a $1\;\text{m}^3$ sensitive volume and $m = 1.1\; \text{MeV}$, this yields $6\cdot10^{-4}$ events per day.\\
Experimentally, the signal would consist in the simultaneous detection of two $511\;\text{keV}$ $\gamma$-rays from the positron annihilation together with
an electromagnetic shower from the decay electron, of an energy peaked around a value depending on $m$. Such a signal should be recognizable in
a detector having an energy threshold of $250\;\text{keV}$. \\

In the KamLand detector, $(1000\;\text{m}^3)$ the signal rate would be $0.6$ event/day for $m = 1.1\;\text{MeV}$  \\ 
The signature should be better identified and the background much reduced in a liquid argon detector. In a $20\;000\;\text{m}^3$ detector of the 
{\it Glacier} type, such as proposed in the frame of the {\it Laguna-LBNO} \cite{Laguna} project, the counting rate becomes 12 events/day for 
$m = 1.1\;\text{MeV}$ and still about $.18$ event/day for $m = 100\;\text{MeV}$

\begin{figure}
\begin{center}
\epsfig{file= 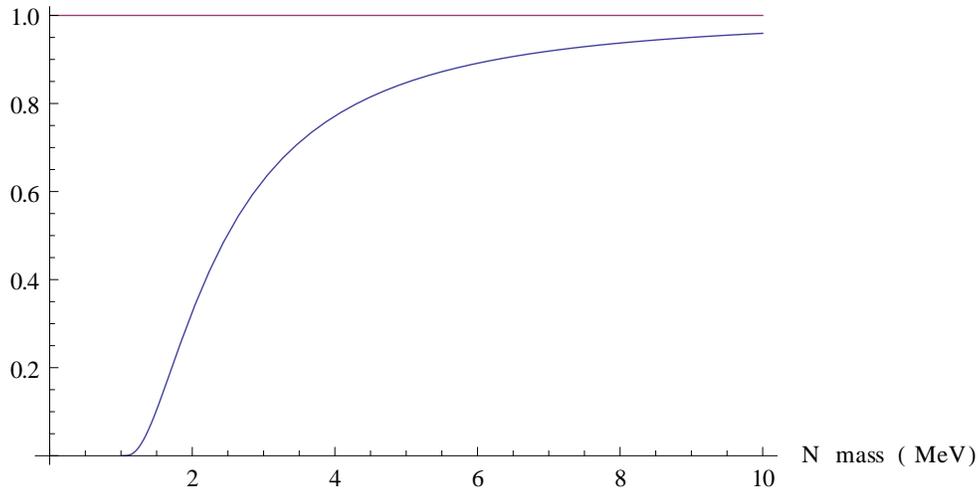,width=13cm}
\caption{The function $f$ which appears in $\Gamma(N)$}
\end{center}
\end{figure}

\end{document}